\documentclass[runningheads]{llncs}
\usepackage{graphicx}
\usepackage{amsmath}
\usepackage{xcolor}

\usepackage{tabu}

\usepackage[utf8]{inputenc}

\usepackage{amsfonts}
\usepackage{amsmath}

\usepackage[parfill]{parskip}

\newcommand{\dotp}[2]{\langle #1, #2 \rangle}

\newcommand\opn{\mathrel{\ooalign{$\subseteq$\cr
  \hidewidth\raise.225ex\hbox{$\circ\mkern.5mu$}\cr}}}

\newcommand{\Smat}{\mathbf{S}}

\newcommand{\us}{\text{\tiny{US}}}
\newcommand{\mr}{\text{\tiny{MR}}}

\newcommand{\qot}[1]{``#1''}

\begin{document}

\title{Global Multi-modal 2D/3D Registration via Local Descriptors Learning}

\author{Viktoria Markova\inst{1, 2}* \and
  Matteo Ronchetti\inst{1}* \and Wolfgang Wein\inst{1} \and Oliver Zettinig\inst{1}  \and Raphael Prevost\inst{1}}

\authorrunning{V. Markova et al.}
 
\institute{ImFusion GmbH, Munich, Germany
\email{info@imfusion.com} \\
\url{https://www.imfusion.com/}
\and Technical University of Munich, Germany }

\renewcommand{\thefootnote}{\fnsymbol{footnote}}
\footnotetext[1]{The two authors contributed  equally to this paper.}

\maketitle              
\begin{abstract}

Multi-modal registration is a required step for many image-guided procedures, especially ultrasound-guided interventions that require anatomical context.
While a number of such registration algorithms are already available, they all require a good initialization to succeed due to the challenging appearance of ultrasound images and the arbitrary coordinate system they are acquired in.
In this paper, we present a novel approach to solve the problem of registration of an ultrasound sweep to a pre-operative image.
We learn dense keypoint descriptors from which we then estimate the registration.
We show that our method overcomes the challenges inherent to registration tasks with freehand ultrasound sweeps, namely, the multi-modality and multidimensionality of the data in addition to lack of precise ground truth and low amounts of training examples.
We derive a registration method that is fast, generic, fully automatic, does not require any initialization and can naturally generate visualizations aiding interpretability and explainability. 
Our approach is evaluated on a clinical dataset of paired MR volumes and ultrasound sequences.

\end{abstract}

\section{Introduction}
Multi-modal registration, the overlaying of data acquired from different modalities, is an important task in the medical computing domain.
A common application is procedure planning and subsequent intervention for diagnostic or treatment. Prostate biopsies, for example, are planned on Magnetic Resonance (MR) images, which may visualize cancerous regions, but executed under ultrasound (US) guidance \cite{nassiri_step-by-step_2019}. Another example is image-guided surgery, which is the standard of care in intracranial neurosurgery \cite{galloway_image-guided_2012} and is also performed for the abdominal section \cite{najmaei_image-guided_2013}.

There exist well-performing local optimization methods for refined registration~\cite{wein_global_2013}, however, they require a good starting pose initialization. In the clinical practise, this is often realized by asking the clinician to select landmarks on both images and confirm a correct registration. This takes time and an active participation of highly trained personal. In addition, the approach is not applicable for automatic batch processing in wide studies. 

In this work, we tackle the registration initialization of US and MRI, in particular of abdominal images. Ultrasound is inexpensive, fast and flexible, but poses multiple challenges: it is quite noisy, contains artifacts and deforms the tissue during acquisition. Moreover, an ultrasound frame has a restricted field of view in an arbitrary orientation contrary to MR or Computed Tomography (CT) images. Additionally, the abdominal area is particularly challenging as the organs and tissues are deformable and highly influenced both by the pressure required for US acquisition and respiration of the patient.  

Until recently, prior works either use manually selected correspondences or a starting pose~\cite{donofrio_abdominal_2019}, a reference tracker or tool that moves the image in the correct vicinity~\cite{favazza_development_2018,donofrio_abdominal_2019,Curious2020}. Another viable automated (but application-specific) approach is registration from segmentation~\cite{thomson_miccai_2020,muller2014}. 

Correspondences between points are essential tool for tackling many computer vision problems such as visual localization~\cite{camera_localization}
simultaneous localization and mapping~\cite{tang_self-supervised_2020} and structure-from-motion~\cite{tang_self-supervised_2020}.
In recent years learned keypoint detectors and descriptors \cite{sun_loftr_2021,dusmanu_d2-net_2019,detone_superpoint_2018,song_sekd_2020} have been successfully applied to computer vision problems involving natural images, replacing hand-crafted approaches such as SIFT \cite{sift} and ORB \cite{orb}.
Keypoint correspondences can be processed to compute robust solutions \cite{ransac}, filtered to enforce constraints \cite{shi_robin_2021} and can provide enough information to estimate non-rigid transformations \cite{deformable_keypoint_registration,dramms}.
Furthermore the visual and intuitive nature of point-to-point correspondences provides interpretability and explainability.
Despite all these advantages, to the best of our knowledge, these approaches have been applied the medical domain only to a limited extend. 
In~\cite{wang_automatic_2017} Wang et al. use keypoints learned by random forests to register mono-modal brain images.
Another paper~\cite{esteban_towards_2019} shows promising registration results from learned keypoints, with the downsides of it being specific to the modalities of X-ray to CT and requiring some manual labeling annotation.

In this paper, we propose a method to discover, extract and match keypoints from ultrasound and MR images.
Our contributions consist in adapting the chosen network method to the multi-modal and cross-dimensional setting (as the ultrasound frames are 2D and the MR image is 3D) while handling imprecise ground-truth.  
In order to not consider the ultrasound geometry but rather the structural features, we further augment the ultrasound data by cropping it in a random polygon shape. 

The method's advantages include its genericity, as it is applicable across modalities and different data dimensions. The method can be easily extended to a deformable one. It is explainable as one can visualize the computed similarities and diagnose for potential problems and biases in the data or training setup.
\section{Approach}
\begin{figure}[t]
    \centering
    \begin{tabular}{c}
    {\includegraphics[width=0.95\linewidth]{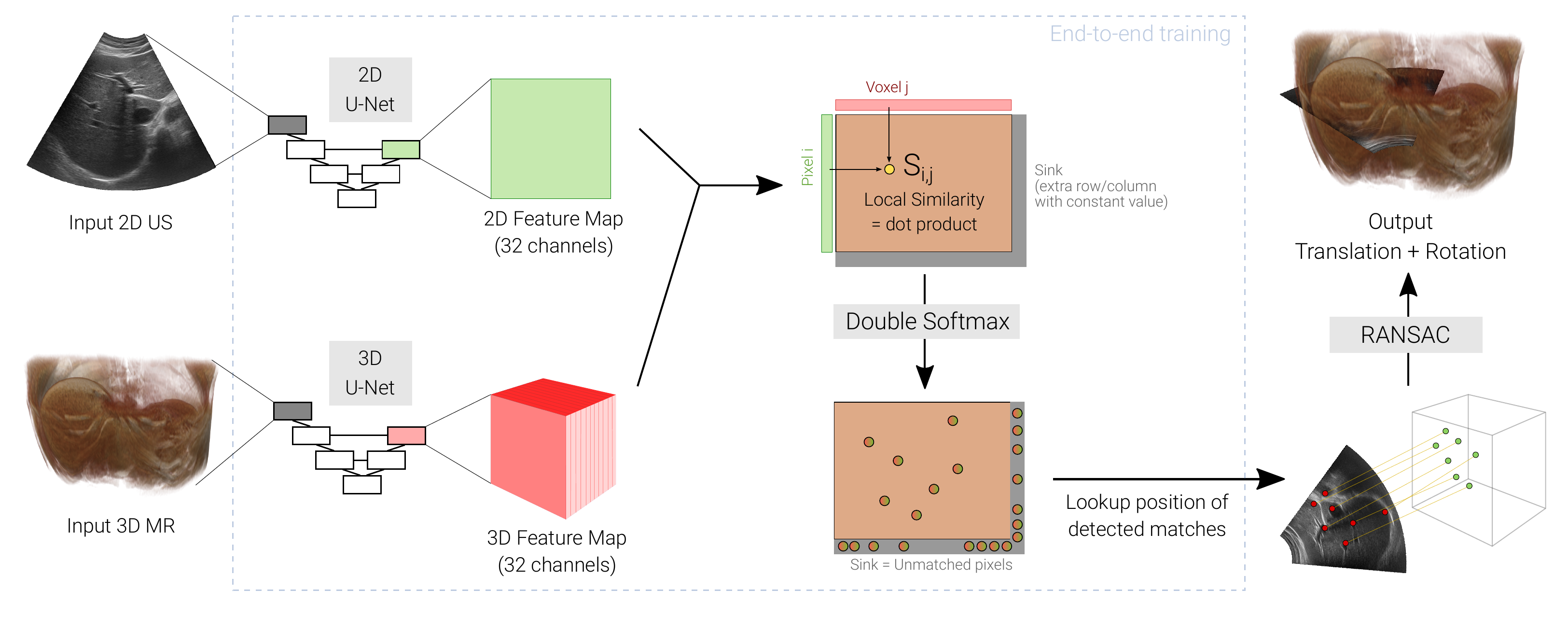}}
     \\
    \end{tabular}
  \caption{Overview of the proposed method. The images are fed to two separate feature extraction networks outputting dense feature maps, which are combined in a matching module, all end-to-end trainable. 
  The matches are extracted via a confidence threshold and processed with RANSAC to retrieve the pose.}
  \label{fig:approach}
\end{figure}

Our registration method, summarized in figure~\ref{fig:approach}, is based on detection and matching of local features across modalities. 
The training is done end-to-end using pose as the only supervision, therefore, it does not require the tedious definition and annotation of specific anatomical landmarks. After extraction of the matches, we classically deploy RANSAC~\cite{ransac} to estimate the pose.

\subsection{Challenges of local feature extraction for medical images}

Our method is based on the LoFTR algorithm proposed in~\cite{sun_loftr_2021}.
There are however a number of key differences between the problem at hand and the original paper (for tracking of 2D natural images), which requires us to generalize and adapt the method in several ways.

\subsubsection{Multi-modality and multi-dimensionality}
Different modalities exhibit different visual appearances and also emphasize different structures.
We overcome this issue by jointly training two distinct networks (a 2D model for US and a 3D one for MR) to produce cross-modality descriptors.

\subsubsection{Inaccuracy of ground truth}
This comes from different sources: MRI and US taken at different times, respiration, deformation caused by probe pressure.
Ground truth registration is generally unreliable to create perfect correspondences. 
As learning a detector on inaccurate correspondences would hardly produce repeatable detections,
we use a detector-free architecture that distributes keypoints uniformly on a grid at $\tfrac{1}{8}$ resolution.
Furthermore, we propose a softer loss that, by taking into consideration the spatial proximity of target keypoints, 
does not over-penalize matches that do not exactly align with the ground truth. 

\subsubsection{Scarcity of data and similarity of motion}
The scarce availability of training data makes it harder to train deep models while avoiding overfitting.
In particular, if the dataset contains sweeps with similar frame geometry and motion, the network might consider the borders of the frame geometry as important spatial clues and take less into consideration the content of the ultrasound frame itself.
We mitigate this issue with data augmentation and design a random masking scheme (PolyCrop) that prevents the overfitting to the frame geometry. 
Moreover, based on early experiments and due to the dataset size, we decided not to use the Transformer and Coarse-to-Fine modules of LoFTR.

\subsection{Detector-free local feature networks}
We use Convolutional Neural Networks (CNN) to extract local features from the US frame and the MR volume.
The architecture of both networks is very similar to a U-Net~\cite{unet} without the last upsampling layers, and makes use of residual blocks~\cite{resnet}, leaky ReLUs~\cite{leaky_relu} non-linearities and instance normalization layers~\cite{instance_normalization}. 
The output of both models is a uniform grid of 32-dimensional descriptors at $\tfrac{1}{8}$ the input resolution.
We indicate the local US and MR features produced by these models with $\mathbf{f}^\us$ and $\mathbf{f}^\mr$ respectively.
A single index $i \in \Omega_\us$ (resp. $j \in \Omega_\mr$) is used to identify a position on the descriptors grid.

\subsubsection{Differentiable matching}\label{sec:matching} 
In order to train the networks end-to-end, the matching too needs to be differentiable.
Similarly to~\cite{sarlin_superglue_2020,sun_loftr_2021}, we define the similarity between two descriptors as their dot product, which allows the networks to encode the quality of a keypoint in the norm of its descriptor.
We thus store the similarity scores in a $|\Omega_\us|\times|\Omega_\mr|$ matrix:
\begin{equation}
\label{eq:similarity}
\Smat_{i,j} = \begin{cases}
\dotp{\mathbf{f}_{i}^\us}{\mathbf{f}_{j}^\mr} & \text{if } i \leq |\Omega_\us| \text{ and } j \leq |\Omega_\mr|\\
\alpha & \text{otherwise}
\end{cases}
\end{equation}
Both the last row and the last column of $\Smat$ are filled with a learned value $\alpha$
which represents a sink for all keypoints that are not matched at all in the other image.
We use the dual-softmax operator~\cite{dual_softmax} to convert the score matrix $\Smat$ into a soft assignment matrix $\mathbf{A}$. Formally, the score of a match $(i, j)$ is defined as:
\begin{equation}
\mathbf{A}_{i,j} = \text{Softmax}_j(\Smat_{i,\cdot}) \cdot \text{Softmax}_i(\Smat_{\cdot,j}) \,. 
\end{equation}

\subsubsection{Loss function}
Ground truth matches are defined by computing soft assignments from the US grid to the MR one.
For every cell $i$ of the US grid, we compute the position of its center $p_i$ and apply the ground truth deformable registration to obtain the corresponding position on the MR $q_i$. We then denote as $m(i)$ the cell corresponding to $q_i$, i.e. the matching cell of $i$.
Our loss is the negative log-likelihood of the ground truth soft matches $w$:
\begin{align}
\mathcal{L} &= - \frac{\sum_{i, j} w(i,j) \log\left(\mathbf{A}_{i,j}\right) }{\sum_{i,j} w(i,j)} \\
\text{where } w(i,j) &= \begin{cases}
\exp(-\beta \| j - m(i) \|)  & \text{if } i \leq |\Omega_\us| \text{ and } j \leq |\Omega_\mr|\\
1 & \text{otherwise}
\end{cases}
\end{align}

\subsection{Multiple frames}
If the ultrasound is acquired with the help of an external tracking system, this can be naturally exploited by the method. The tracking provides relative pose of the frames to each other, which can be used to take into account matches from multiple frames for the final pose estimation with RANSAC~\cite{ransac}. 
\section{Experiments}

\subsection{Datasets}\label{sec:data}

\label{dataset_clinical}
The dataset consists of 16 patients who have each a T1-weighted MR image taken and a number of ultrasound sweeps before surgery. No special sonographic protocol has been enforced during the acquisition, the operator simply acquired one or two sweeps to maximize tumor and liver gland visibility. The ground truth pose has been obtained by manually registering every sweep to its pre-operative MR image. We generate correspondences using defomable registration and evaluate on rigid pose estimation. In total our dataset contains 30 sweeps with 3957 frames. 
We train four-fold cross-validation
and show ablation studies on different settings. 

We resample both MR and ultrasound images to have uniform spacing of 1.25mm.
Sweeps with transversal orientation appear three times more often than intracoastal ones, we therefore sample more frequently the intracostals to balance the dataset. 
Gaussian noise augmentation and random cropping are used to mitigate over-fitting.
Finally, since we noticed a particular sensitivity of the networks to the ultrasound field of view, we mask (i.e. set the pixels to zero) the outer part of each ultrasound image with a random convex polygon (\qot{PolyCrop}).
This ensures that the features to be extracted use actual image information instead of exploiting the ultrasound frame geometry. 

\subsection{Baselines and main results}

Our baseline for comparison is the method proposed in~\cite{muller2014}, which is based global registration of segmentation maps.
We make use of the proprietary implementation available in the ImFusion Suite 2.36.3 (ImFusion GmbH, Munich, Germany) software, which segments liver, diaphragm and vena cava.

We report the results on the pose error with mean, standard deviation and median (in degrees for the rotation part, mm for the translational part). Additionally, we comment on the statistical significance of the results. 

Our registration method is evaluated both as-is (\qot{Initialization}), and as an initialization for a local multi-modal registration algorithm (\qot{After Registration}).
This algorithm, which optimizes the pose of the US volume together with the deformation of the MR, has been shown to achieve good results~\cite{wein_global_2013} but requires a close initialization. 
While this local registration is not part of our work, it is relevant to study whether our pose initialization falls into the capture range of such an algorithm.
For those experiments, we used the implementation provided in the same software.
Table~\ref{tab:results1} shows the comparison between the proposed method and the baseline.
Our approach is evaluated both using a single US frame (no tracking required) and using the entire sweep.
Even on single frames, our method significantly outperforms the baseline.
The superiority is statistically significant as shown by the Wilcoxon signed-rank test. The p-values for all four reported metrics are $\leq10^{-5}$ for the multi-frame variant (as this is better comparable to the baseline method).

\begin{figure}[t]
    
    \includegraphics[width=\linewidth]{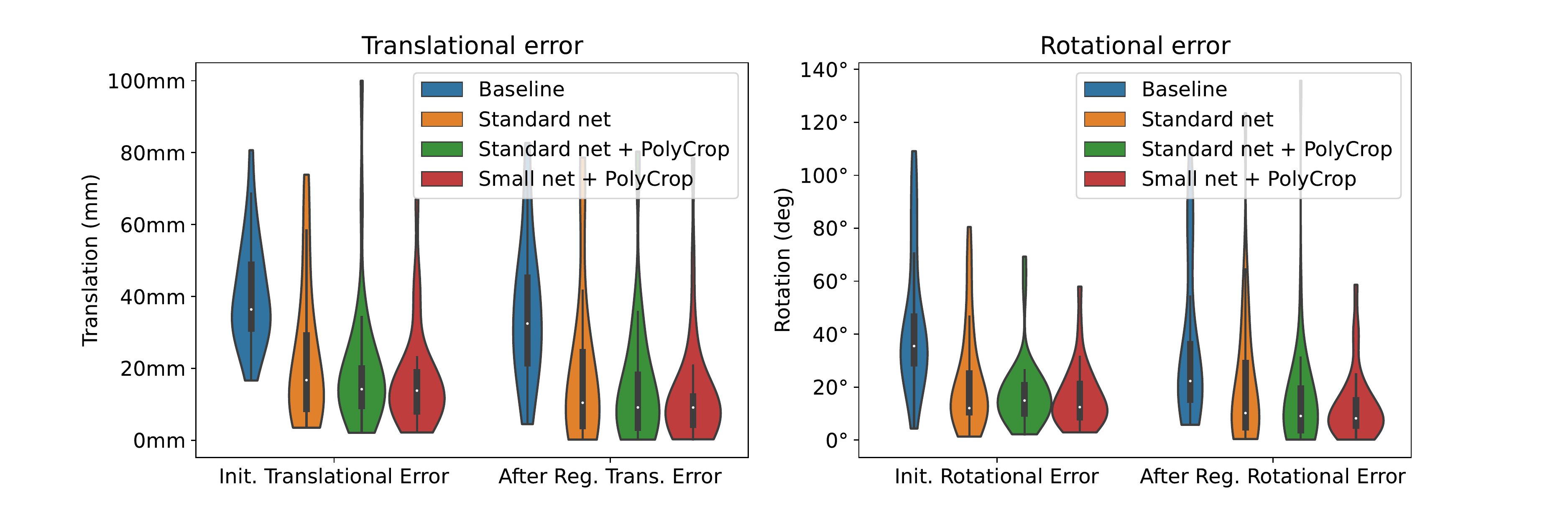}
    \caption{Violin plots of the distributions of translational (left) and rotational (right) errors split across methods. Black rectangles are boxplots, white dots are medians.
    }
  \label{fig:plot}
\end{figure}

The last two rows of the table quantify the progress towards the ultimate goal of fully automatic multi-modal registration.
The threshold of the fifth row is chosen in accordance with the reported capture range of the local optimization algorithm~\cite{wein_global_2013}.
Our method favors relatively well with 45\% with tracking and 32\% without. The percentages grow significantly with the broader threshold depicted in the row below.
The distribution of pose errors is depicted in Figure~\ref{fig:plot} using a violin plot.
It can be noticed that, while the majority of cases are registered with good results, a few failure cases exist.

\begin{table}[ht]
\caption{Comparison between pose prediction error between the proposed method, both with tracking (multiple frames) and without (single frame) and the baseline.}
\label{tab:results1}
\scriptsize
\centering
    
    \renewcommand\arraystretch{1.25}
    \setlength{\tabcolsep}{1mm}
    {\begin{tabular}{l|c|cc}
     &
    Baseline &
    \multicolumn{2}{c}{Proposed}
    \\
     &
    All frames &
    All frames &
    Single frame
    \\
    \hline
    
    Init. Rot. (deg) & 
    \(44.0 \pm 26.2 (35.5)\) &
    \( 16.2 \pm 11.9 (12.4) \) &
    \( 21.3 \pm 16.3 (18.4) \)
    \\
    
    Init. Trans. (mm) & 
    \(40.5 \pm 14.8 (14.4)\) &
    \( 16.7 \pm 13.8 (13.8) \) &
    \( 21.5 \pm 16.9 (17.8) \)
    \\
    \hline
    
    After Reg. Rot. (deg)  & 
    \(34.6 \pm 29.0 (22.3)\)  &
    \( 12.7 \pm  12.9 (8.2) \) &
    \( 17.5 \pm 18.4 (13.1)\)
    \\
    
    After Reg. Trans. (mm) & 
    \(35.5 \pm 19.6 (32.5)\) &
    \( 12.7 \pm  15.9 (9.1) \) &
    \( 17.4 \pm 18.0 (10.9) \)
    \\
    \hline

    After R. $<10$\textdegree{} \& $<10$mm & 
    6.7 \% &
    45.2 \% &
    32.3 \%
    \\
    
    After R. $<15$\textdegree{} \& $<20$mm  & 
    10.0 \% &
    74.2 \% &
    49.5 \%
    \\
    \end{tabular}
    }
\end{table}

\begin{table}[ht]
    \caption{Ablation studies results. Comparison between pose prediction errors. In all cases the tracking has been utilized 
    and matches from multiple frames used for the pose estimation.
    }
    \label{tab:ablation_studies}
\scriptsize
\centering
    \renewcommand\arraystretch{1.25}
    \setlength{\tabcolsep}{1mm}
    {\begin{tabular}{l|c|c|c}

    \hfill \bfseries Method \quad &

    No polycrop &
    Polycrop &
    Polycrop 
    \\ \bfseries Metric &
    + standard net &
    + standard net &
    + smaller net 
    \\
    \hline
    
    Init. Rotation (deg)  & 
    \(22.4 \pm 19.5 (12.0)\)  &
    \(18.0 \pm 14.3 (14.7)\) &
    \( 16.2 \pm 11.9 (12.4) \)
    \\
    
    Init. Trans. (mm) & 
    \(23.1 \pm 19.6 (16.7)\)  &
    \(19.9 \pm 20.0 (14.2)\) &
    \( 16.7 \pm 13.8 (13.8) \)
    \\
    \hline
    
    After Reg. Rotation (deg)  & 
    \(21.0 \pm 26.0 (10.2)\)  &
    \(15.7 \pm 25.7 (6.5)\) &
    \( 12.7 \pm  12.9 (8.2) \) 
    \\
    
    After Reg. Trans. (mm) & 
    \(19.5 \pm 22.7 ({10.4})\)  &
    \(15.3 \pm 20.1 (8.8)\) &
    \( 12.7 \pm  15.9 (9.1) \) 
    \\
    \hline
    
    Init. $>$50\textdegree{} or $>$50mm & 
    \(12.9 \%\)  &
    \(6.7 \%\) &
    3.2 \% 
    \\
    
    Init. $<$25\textdegree{} \& $<$25mm  & 
    \( 58.1\%\)  &
    \(73.3\% \) &
    80.6\% 
    
    \\

    \end{tabular}
    }
\end{table}

\subsection{Ablation studies}
We quantify the effects of Polycrop augmentation and network size through ablation studies whose results are summarized in Table~\ref{tab:ablation_studies}.
For easier interpretation, the fifth and sixth rows show respectively the percentage of high-error and well-initialized cases.
The error distributions are also depicted in Figure~\ref{fig:plot} with violin plots.

The column \qot{Polycrop + standard net} (Table~\ref{tab:ablation_studies}) contains the results with the PolyCrop augmentation described in section~\ref{sec:data}. We observe an improvement in the pose initialization compared to \qot{No poly + standard net}. Both the median in translation and rotation error in the initial pose drop by more than~3 degrees and millimeters respectively. These improvements, naturally, carry over to the end pose after the local optimization.

We experiment with a smaller version of our networks and compare \qot{Polycrop + standard net} with \qot{Polycrop + smaller net}.
A size reduction in the network leads to faster training and inference, as well as, smaller memory requirements.
The smaller model achieves better results on all the metrics except on the median error after registration refinement.
A possible cause of this improvement is that the smaller model, being less subject to over-fitting, produces a lower number of high-error cases while being slightly worse in the majority of cases. This can be observed in Figure~\ref{fig:plot} by noticing the shorter tails of the distribution.

\subsection{Similarity visualization}
\begin{figure}[t]
    \centering
    \begin{tabular}{cc}
     {\includegraphics[width=0.40\linewidth, height=3.2cm]{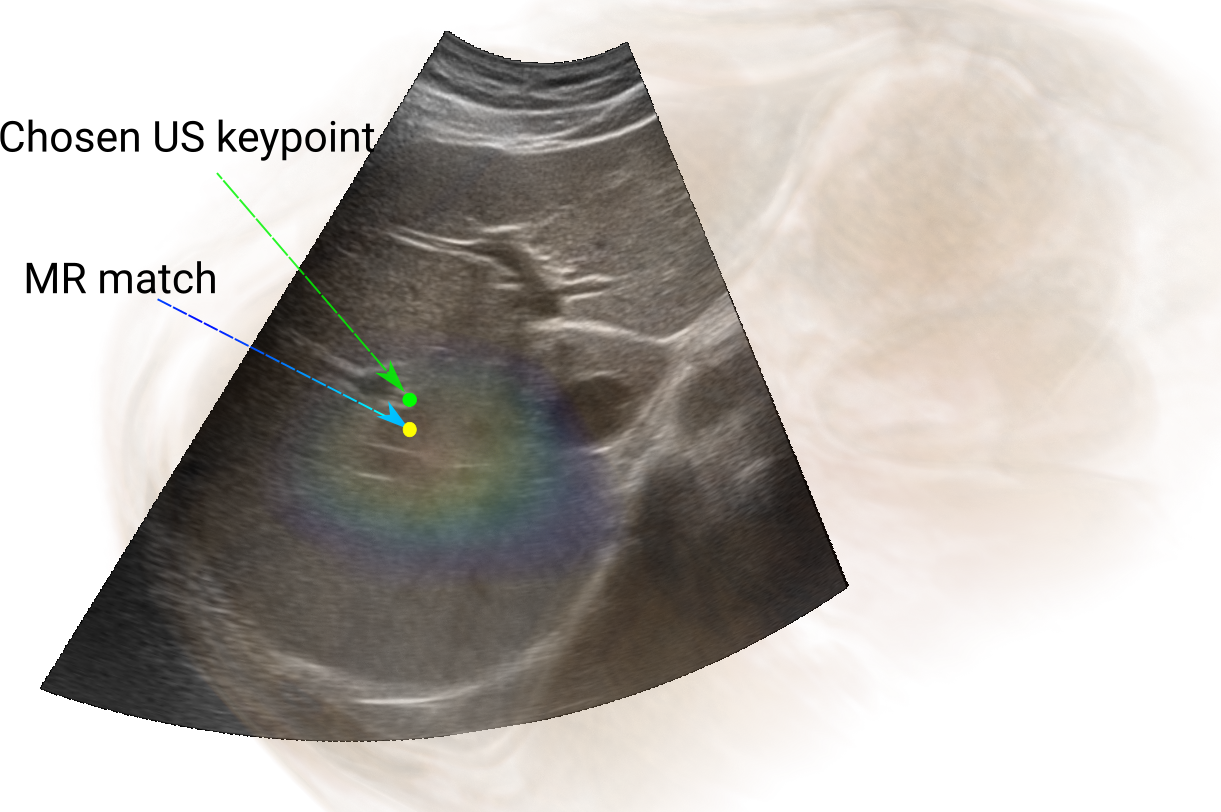}}
     &
    {\includegraphics[width=0.30\linewidth, height=3.2cm]{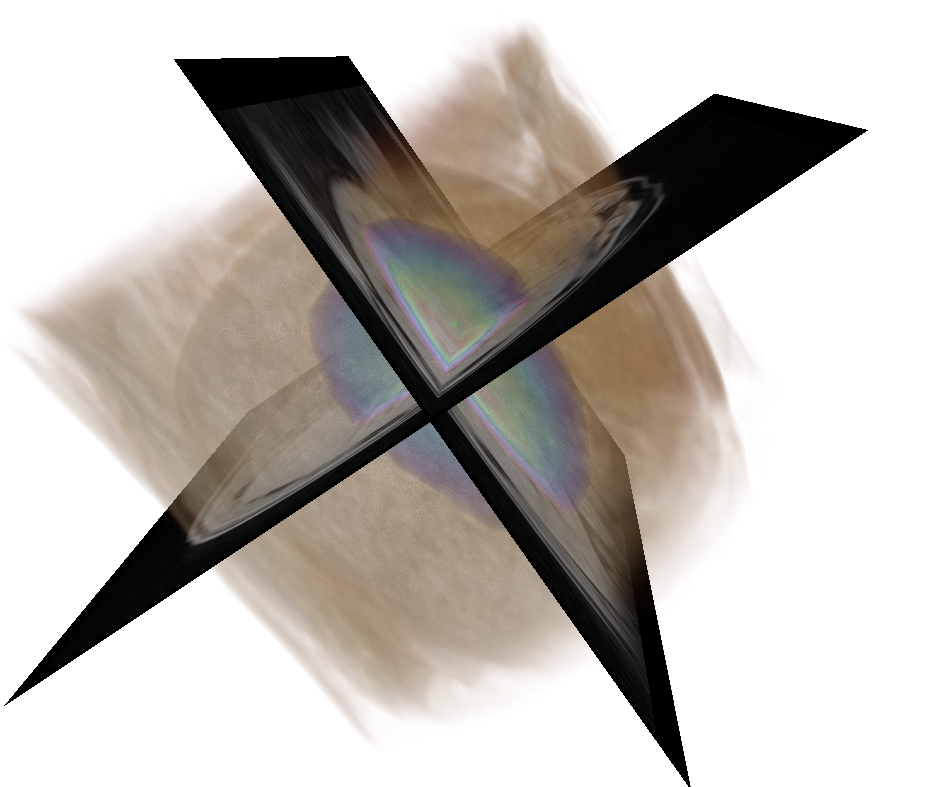}}
    {\includegraphics[width=0.25\linewidth, height=3.2cm]{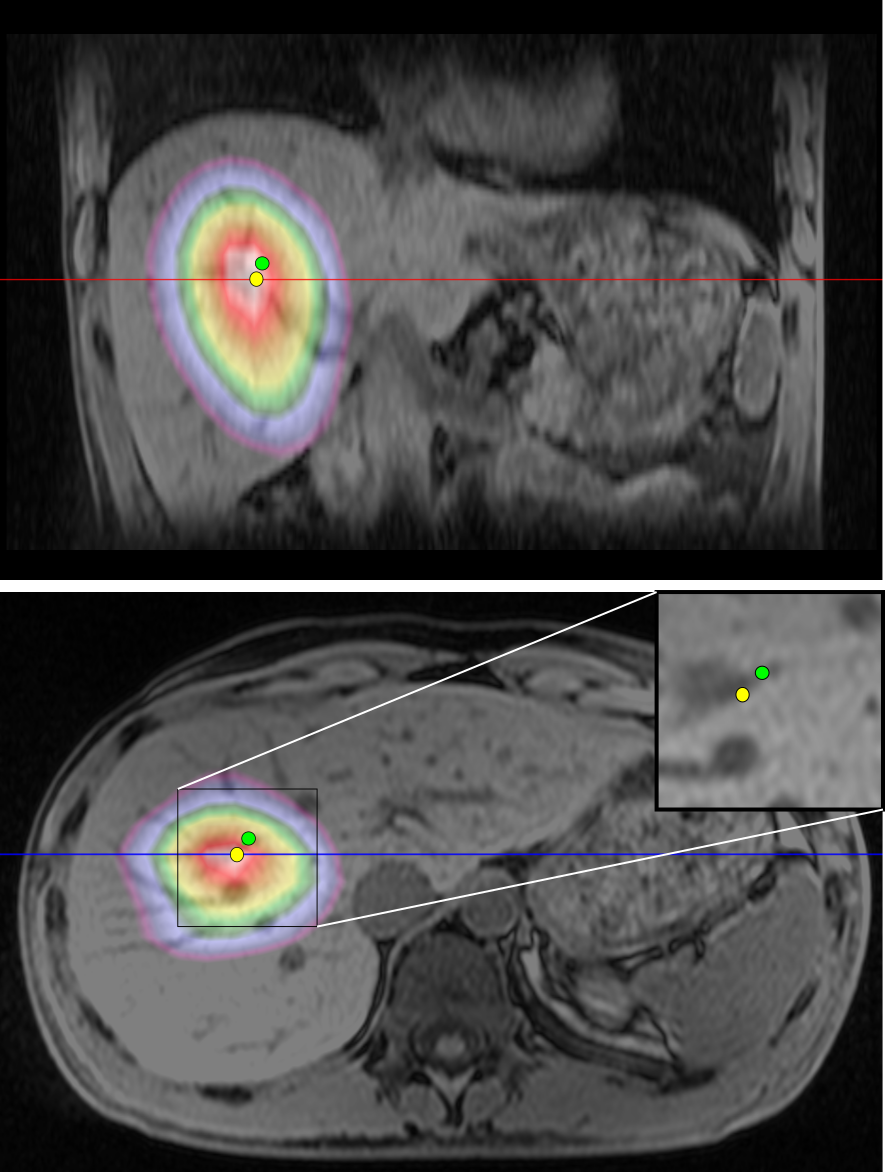}}
     \\
     (a) US frame & (b) MR intersections
    \end{tabular}
  \caption{Similarity visualization. On the left (a), the US frame with the chosen keypoint, it's corresponding match on the MR. The MR and US are registered with the ground truth, thus if I the method were perfect, the two points would be at the same position. On the right (b) two intersections of the MR and their visualization.}
  \label{fig:heatmap_mri}
\end{figure}

Our method can be inspected and its results visualized and interpreted. Figure~\ref{fig:heatmap_mri} depicts, as a heatmap, a row of the similarity matrix  $ \mathbf{S} $ (defined in Eq.~\ref{eq:similarity}) i.e. the similarity between each MR voxel and the selected point in the US frame (green dot).
The matched MR keypoint (maximum of the heatmap) is shown as a yellow dot.
The shape of the heatmap can be used to identify the main direction of uncertainty, in this particular example its elongation across the axis of motion corresponds to the presence of tubular features.

\textbf{}
\section{Conclusions}

In this paper we proposed a new approach for multi-modal image registration.
Our method, based on learned keypoint detection and matching, is fast, fully-automatic, interpretable and can be generalized to other image modalities.
Our approach generalizes and adapts one of the latest advances from the computer vision community to overcome challenges inherent to multi-modal and multi-dimensional medical images.

On challenging abdominal US + MR data, our method can fully automatically pre-align the data in $>$80\% of all attempts for tracked 3D freehand sweeps. In addition, a single US frame successfully registers in about half of the attempts; this can yield significant workflow improvements for interventional scenarios where time is of the essence, and hence the ultrasound registration step is required to be brief and simple. We, therefore, believe that the proposed method can become an important ingredient in image-guided surgery systems that require intra-operative registration, as well as easier to use diagnostic fusion systems.

This study focused on rigid pose estimation in the context of registration initialization. 
A natural extension of this work, since the method produces more matches than necessary, would be to use the learned correspondences to estimate a dense deformation field.
The single frame performance of our algorithm suggests an application in Ultrasound SLAM, where the sweep motion is unknown and needs to be reconstructed.
Capture range of image based registration methods, might be improved by using the dense local descriptors produced by our networks to define a distance metric between images.
Finally, it should be possible to improve on our results by tuning the network architecture and collecting a larger dataset or incorporating other supervisory signals, such as segmentation, into the training loss.

\newpage

\bibliographystyle{splncs04}

\bibliography{citations}

\end{document}